\documentclass[twocolumn]{aastex62}

\usepackage{comment}

\graphicspath{{./}{figures/}}

\received{December 31, 2018}
\revised{March 19, 2019}
\accepted{March 28, 2019}
\submitjournal{ApJL}

\shorttitle{Breezing through the space environment of Barnard's Star b} 
\shortauthors{Alvarado-G\'omez \& Garraffo et al.}

\begin{document}

\title{Breezing through the space environment of Barnard's Star b}

\correspondingauthor{Juli\'an D. Alvarado-G\'omez, Cecilia Garraffo}
\email{jalvarad@cfa.harvard.edu, cgarraffo@cfa.harvard.edu}

\author[0000-0001-5052-3473]{Juli\'an D. Alvarado-G\'omez}
\altaffiliation{These authors contributed equally to this work.}
\affil{Center for Astrophysics $|$ Harvard \& Smithsonian, 60 Garden Street, Cambridge, MA 02138, USA}

\author[0000-0002-8791-6286]{Cecilia Garraffo}
\altaffiliation{These authors contributed equally to this work.}
\affil{Institute for Applied Computational Science, Harvard University, Cambridge, MA 02138, USA}
\affil{Center for Astrophysics $|$ Harvard \& Smithsonian, 60 Garden Street, Cambridge, MA 02138, USA}
\collaboration{}

\author{Jeremy J. Drake}
\affil{Center for Astrophysics $|$ Harvard \& Smithsonian, 60 Garden Street, Cambridge, MA 02138, USA}

\author{Benjamin P. Brown}
\affil{Department of Astrophysical and Planetary Sciences, Laboratory for Atmospheric and Space Physics, University of Colorado at Boulder 3665 Discovery Drive, Boulder, CO 80303, USA}

\author{Jeffrey S. Oishi}
\affil{Department of Physics and Astronomy Bates College, Carnegie Science Hall, 2 Andrews Road, Lewiston, ME, 04240, USA}

\author{Sofia P. Moschou}
\affil{Center for Astrophysics $|$ Harvard \& Smithsonian, 60 Garden Street, Cambridge, MA 02138, USA}

\author{Ofer Cohen}
\affil{University of Massachusetts at Lowell, Department of Physics \& Applied Physics, 600 Suffolk Street, Lowell, MA 01854, USA}



\begin{abstract}
\noindent A physically realistic stellar wind model based on Alfv\'en wave dissipation has been used to simulate the wind from Barnard's Star and to estimate the conditions at the location of its recently discovered planetary companion. Such models require knowledge of the stellar surface magnetic field that is currently unknown for Barnard's Star. We circumvent this by considering the observed field distributions of three different stars that constitute admissible magnetic proxies of this object. Under these considerations, Barnard's Star b experiences less intense wind pressure than the much more close-in planet Proxima~b and the planets of the TRAPPIST-1 system. The milder wind conditions are more a result of its much greater orbital distance rather than in differences in the surface magnetic field strengths of the host stars. The dynamic pressure experienced by the planet is comparable to present-day Earth values, but it can undergo variations by factors of several during current sheet crossings in each orbit. The magnetospause standoff distance would be $\sim$\,$20 - 40$\,\% smaller than that of the Earth for an equivalent planetary magnetic field strength. 
\end{abstract}



\setcounter{footnote}{3}

\section{Introduction} \label{sec:intro}

\noindent The detection by \citet{2018Natur.563..365R} of a planet around Barnard's Star is an important step in our growing understanding of the nature of planetary systems in the Universe. The M3~V red dwarf Barnard's Star is the closest 
single star planetary system to the Sun. 

Barnard's Star b (BSb) orbits at a distance similar to that of Mercury around the Sun. \citet{2018Natur.563..365R} note that the planet resides close to the ``snow line" of Barnard's Star, where stellar irradiation is sufficiently weak to allow volatile elements to condense \citep{2008ApJ...673..502K}. This characteristic renders BSb of special interest from the perspective of planet formation. 
There is growing agreement on the importance of the snow line region as a natal site of planetesimal formation and growth (e.g., \citealt{1988Icar...75..146S, 2015Icar..258..418M, 2015ApJ...814..130M}) and BSb promises to be an important keystone object for future progress.

Of more immediate interest is the potential of BSb for understanding planetary atmospheric evolution.
It has an orbital semi-major axis of $\sim$\,$0.4$ AU, which is sufficiently wide that it can be disentangled from the stellar signal and will be amenable to detailed study and direct imaging by next generation instrumentation \citep{2016JATIS...2a1013T}. This will allow the planetary atmosphere to be studied in unprecedented detail.

Several papers have pointed to the importance of extreme stellar activity and winds for understanding the evolution of planetary atmospheres around M dwarf stars (e.g., \citealt{2009A&A...506..399L, 2014ApJ...790...57C, 2015MNRAS.449.4117V, 2017ApJ...843L..33G}). Magnetospheric and atmospheric evolution models are not yet able to predict how atmospheric initial conditions evolve under intense radiation environments. Nevertheless, progress has been made in this direction through 1D models of evaporating exoplanet atmospheres and stellar winds (Johnstone et al. \citeyear{2015A&A...577A..28J,2015A&A...577A..27J,2015ApJ...815L..12J}). Based on detailed magnetohydrodynamic (MHD) modeling of the wind of Proxima Centauri, \citet{2016ApJ...833L...4G} found that Proxima~b is subject to dynamic wind pressures four orders of magnitude larger than the Earth.    
The relatively long rotation period of Barnard's star of $\sim$\,$130 - 145$ days \citep{1998AJ....116..429B}, combined with its modest present-day magnetic activity level \citep{2018arXiv181206712T}, lead \citet{2018Natur.563..365R} to estimate an age of 7--10~Gyr for the system. BSb represents the outcome of planetary evolution over this long timescale through a history of strongly differing environmental conditions.

Here, we use detailed 3D MHD stellar wind models to examine the space weather conditions of the present-day Barnard's Star environment and investigate the influence of the stellar wind on its magnetosphere and atmosphere. Our wind models are driven by the observed magnetic field maps of three proxy stars, scaled to the expected range of surface field strength for Barnard's Star.

\section{Methods} \label{sec:Methods}

\subsection{Corona and Stellar Wind}\label{sec:Wind}

\noindent To simulate the environment in the Barnard's Star system, we employ the \href{http://csem.engin.umich.edu/tools/swmf/}{Space Weather Modeling Framework} (SWMF, see \citealt{2018LRSP...15....4G}). In particular, we use the 3D MHD solver BATS-R-US \citep{2012JCoPh.231..870T}, and the Alfv\'en Wave Solar Model (AWSoM, \citealt{2014ApJ...782...81V}). This model was developed to study the solar 
environment and has since been adapted and applied to astrophysical systems. It uses the distribution of the radial magnetic field on the surface of the star (magnetogram) as a boundary condition for the self-consistent calculation of the coronal heating and stellar wind acceleration due to Alfv\'en wave turbulent dissipation. 

The AWSoM model assumes a non-ideal MHD regime, where the magnetically-driven contributions are included as additional source terms in the energy and momentum equations, solved alongside the mass conservation and magnetic induction equations on a spherical grid. Radiative losses and electron heat conduction are also taken into account in the model. The simulation evolves until a steady-state solution is reached. Further details on the implementation and recent performance updates can be found in \cite{2014ApJ...782...81V} and \cite{2016arXiv160904379S}.

In previous studies we have used this code to simulate the space weather conditions around other M~dwarf planet-hosting stars (cf. Prox Cen: \citealt{2016ApJ...833L...4G}, TRAPPIST-1: \citealt{2017ApJ...843L..33G}). Unlike the Prox Cen or TRAPPIST-1 planets, the BSb orbit, with a semi-major axis of $a = 0.404 \pm 0.018$ AU ($\sim$\,$443.16 \pm 19.74$~R$_{*}$) and a relatively high eccentricity ($e = 0.32^{+0.10}_{-0.15}$), resides well outside the domain of the stellar corona. We 
use two coupled numerical domains which include the corona and wind acceleration region (the Solar Corona, or SC,  module of the SWMF; $\sim$\,$1 - 110$~R$_{*}$), and a wind propagation region (the Inner Heliosphere, or IH module; $105 - 750$~R$_{*}$) sufficiently large to enclose the maximum possible orbital separation of the planet. The non-ideal MHD effects described before are only considered in the SC domain, whose solution is then propagated into the IH ideal MHD scheme. A $5$~R$_{*}$ domain overlap is imposed to guarantee the robustness of the combined solution. 

The IH domain utilizes 
adaptive mesh refinement for increased spatial discretization of regions with strong gradients in wind density occurring along the current sheet. In this domain, our simulations reach a smallest cell size of $\sim$\,$1.46$~R$_{*}$ and consider more than $25$ million spatial blocks. In the stellar context, this coupled modeling scheme has been employed to investigate the stellar wind properties in the habitable zones of Sun-like stars (cf.~Alvarado-G\'omez~et~al.~\citeyear{2016A&A...588A..28A}, \citeyear{2016A&A...594A..95A}), and this particular study represents its first application to the M~dwarf regime.  

\subsection{Surface Magnetic Field}\label{sec:Dynamo}

\begin{figure*}[!t]
\centering
\includegraphics[trim=0.2cm 0.2cm 0.2cm 0.2cm, clip=true,width=0.331\textwidth]{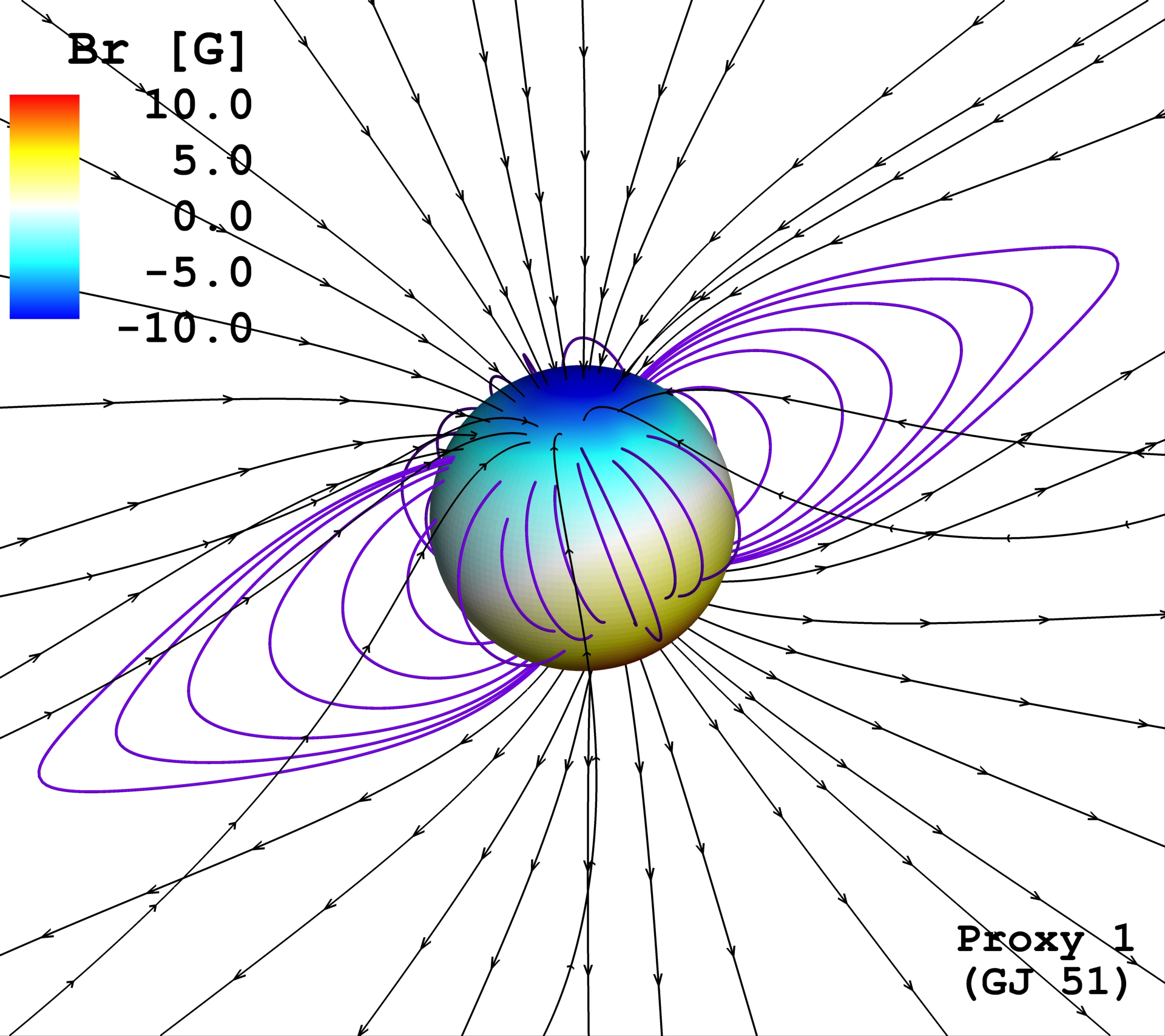}\hspace{1.2pt}\includegraphics[trim=0.2cm 0.2cm 0.2cm 0.2cm, clip=true,width=0.331\textwidth]{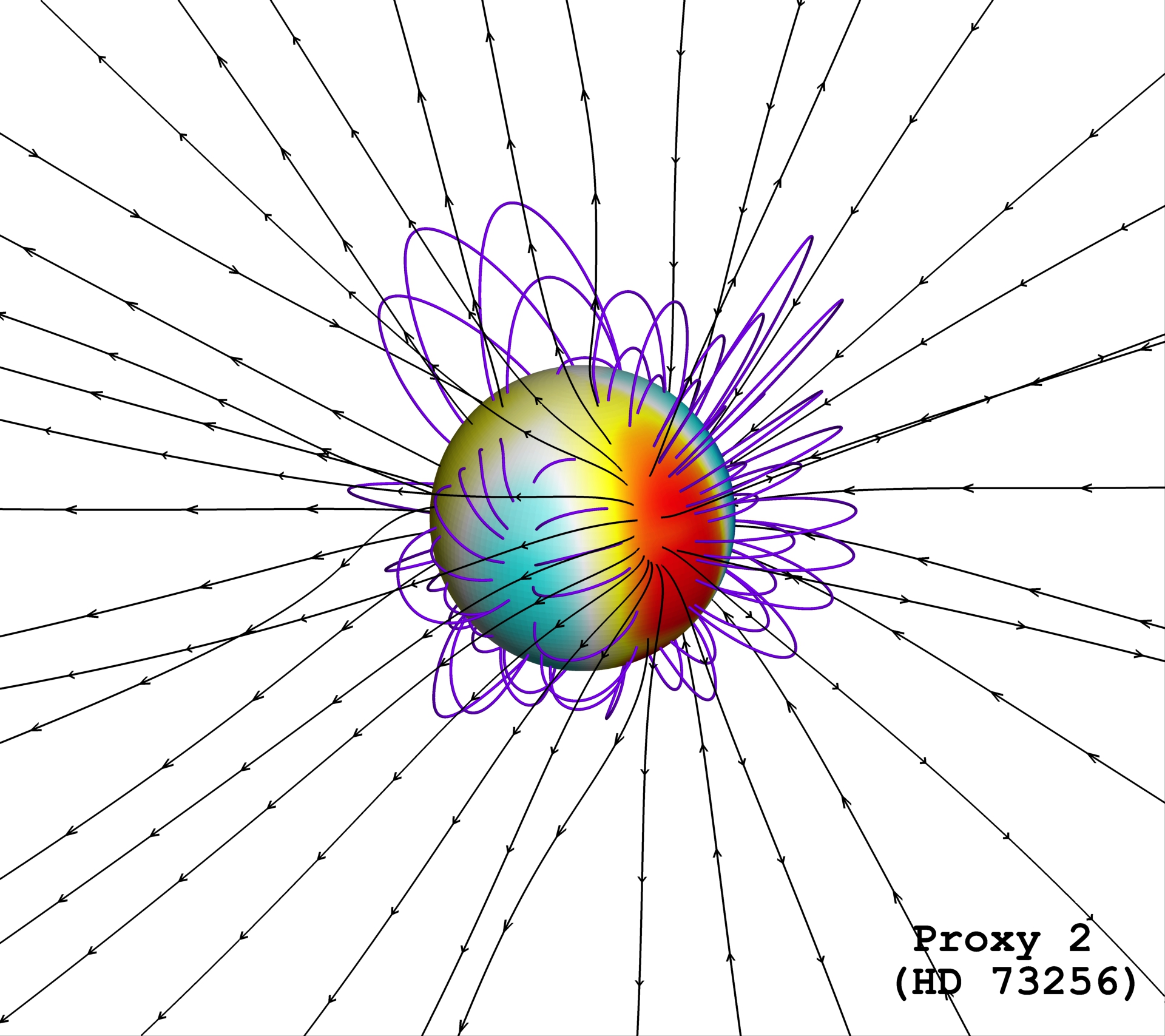}\hspace{1.2pt}\includegraphics[trim=0.2cm 0.2cm 0.2cm 0.2cm, clip=true,width=0.331\textwidth]{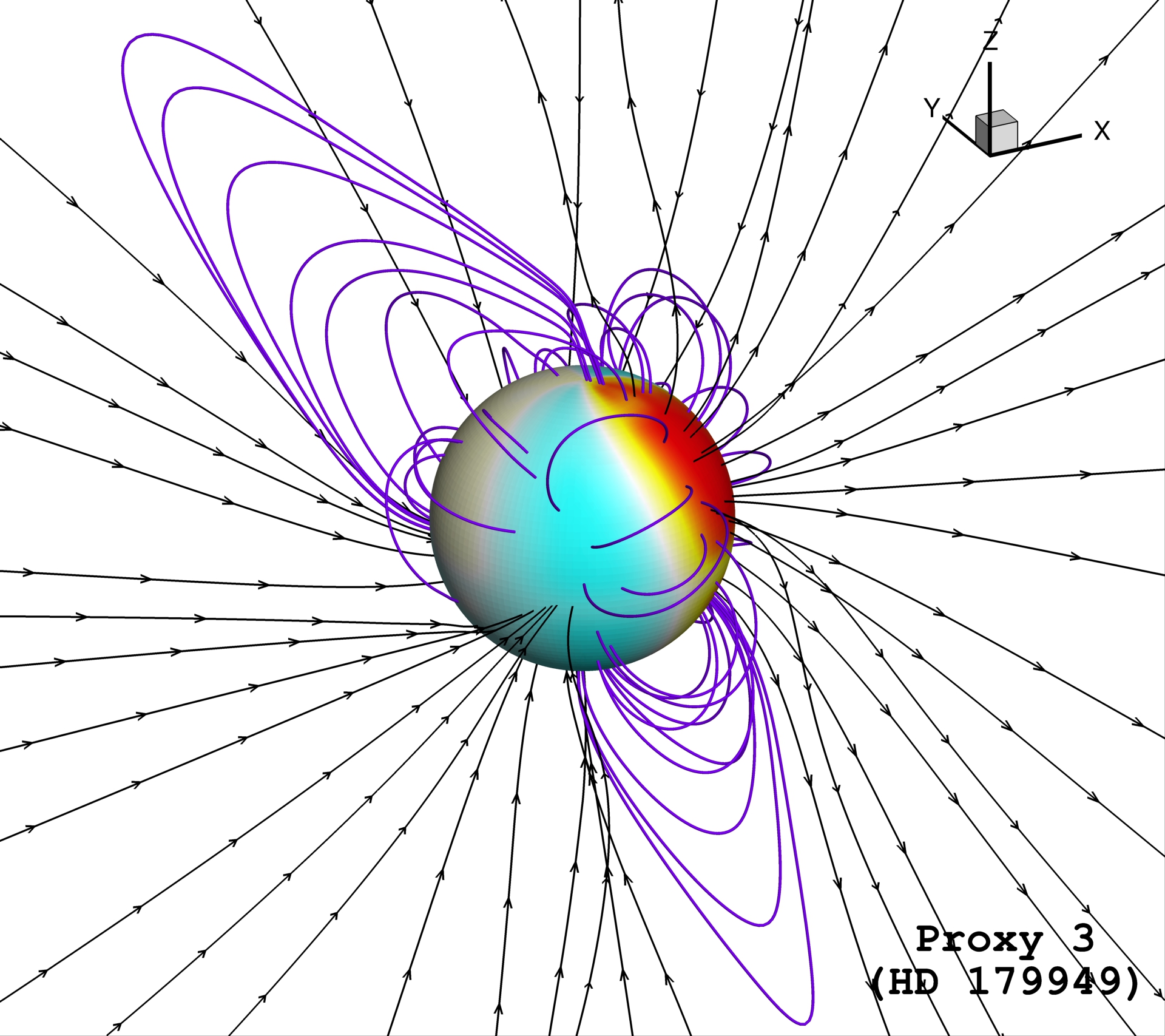}
\caption{Inner boundary of our simulation domain, showing the large-scale ZDI radial magnetic field of three late-type stars driving the proxy models of Barnard's Star (Left:~GJ~51, \citealt{2010MNRAS.407.2269M}; Middle: HD~73256, \citealt{2013MNRAS.435.1451F}; Right: HD~179949, \citealt{2012MNRAS.423.1006F}). The field strengths have been scaled to the indicated range. Selected closed (purple) and open (black) magnetic field lines are shown.}
\label{fig_1}
\end{figure*}

\noindent While high-resolution magnetograms are readily available for the Sun (e.g.\ from the \href{https://sohowww.nascom.nasa.gov/}{Solar and Heliospheric Observatory} and the \href{https://sdo.gsfc.nasa.gov}{Solar Dynamics Observatory}), for stars the only source of magnetograms at present is through spectropolarimetric observations and Zeeman-Doppler Imaging (ZDI), which requires a minimum brightness and rotation velocity \citep{1989A&A...225..456S, 1997MNRAS.291..658D}. Unfortunately, with a visual magnitude of $V_{\rm mag} = 9.511$ and a rotation period of $\sim$\,$130$ days, the retrieval of a ZDI magnetic map of Barnard's Star is currently unfeasible. 
As described by \citet{2012LRSP....9....1R}, measurements in unpolarized light of the unsigned surface magnetic field strength via Zeeman broadening benefit from small rotational broadening.  However, to our knowledge, there are no magnetic field measurements in the literature for Barnard's Star using this technique.

Despite the lack of direct measurements, we can use information from previous studies to get a sense of the field strength and geometry expected for Barnard's Star. On one hand, X-ray luminosity is known to be strongly correlated with stellar rotation, and even more so with Rossby number ($Ro$), defined as the rotation period ($P_{rot}$) over the convective turnover time ($\tau$). \cite{Wright.etal:11} used a sample of 725 Sun-like and later-type stars to calibrate this relationship. Interestingly, in two follow up papers \cite{Wright.Drake:16, Wright.etal:18} showed that this relationship also holds for fully convective stars like Barnard's Star. Making use of this calibration, and the level of X-ray emission of Barnard's Star ($\log L_{\rm X} = 25.85$, \citealt{1999A&AS..135..319H, 2004A&A...417..651S}), we estimate $\tau \sim 120-150$ days, which results in $Ro \sim 0.9-1.2$.

Additionally, detailed numerical MHD convection models appropriate for fully-convective stars have demonstrated that these objects are able to generate global-scale magnetic fields in their convection zones, despite the lack of a solar-like tachocline. In these simulations, especially at low $Ro$, significant magnetic energy has been found in the dipolar components of the field \citep{Browning_2008, Yadav_et_al_2015}. At higher $Ro$, the mean fields can continue to show strong dipole fractions even as the fluctuating fields increase in strength. In addition, spin-down models predict dipole-dominated morphologies in these large Ro regimes (see \citealt{Garraffo.etal:18}). In this context, we consider as one proxy for  Barnard's Star, the ZDI map of the star GJ~51 \citep{2010MNRAS.407.2269M}, which has a dipole-dominated geometry and similar spectral type (M5), albeit with a significantly shorter rotation period ($\sim$\,$1$ day). This is the same magnetogram used in the study of the space weather of Proxima Cen by \citet{2016ApJ...833L...4G} and allows us to perform a comparative analysis of the wind conditions around Proxima and Barnard's Star.

Furthermore, from the compilation gathered by \citet{Vidotto.etal:14}, there are eight stars with $Ro >~0.9$, from which only four have published ZDI magnetic field maps: HD~78366 ($Ro \gtrsim 2.78$, \citealt{2011AN....332..866M}), HD~146233 ($Ro = 1.32$, \citealt{2008MNRAS.388...80P}), HD~73256 ($Ro = 0.96$, \citealt{2013MNRAS.435.1451F}), and HD~179949 ($Ro \gtrsim~1.72$, \citealt{2012MNRAS.423.1006F}). Both, HD~73256 and HD~146233, are close to the estimated $Ro$ values for Barnard's Star and could serve as additional proxies for its field geometry. However, the resolution of the published ZDI map for the latter is too low for our numerical purposes. Therefore, we consider instead the map of HD~179949, which actually shows substantial similitude to the large-scale field of HD~146233. All maps considered here were taken directly from the publications (via an image-data transformation), and therefore they retain the intrinsic missing latitudes of ZDI observations. 

\begin{figure*}[!ht]
\centering
\includegraphics[trim=0.15cm 7.0cm 0.15cm 7.5cm, clip=true,scale=0.16]{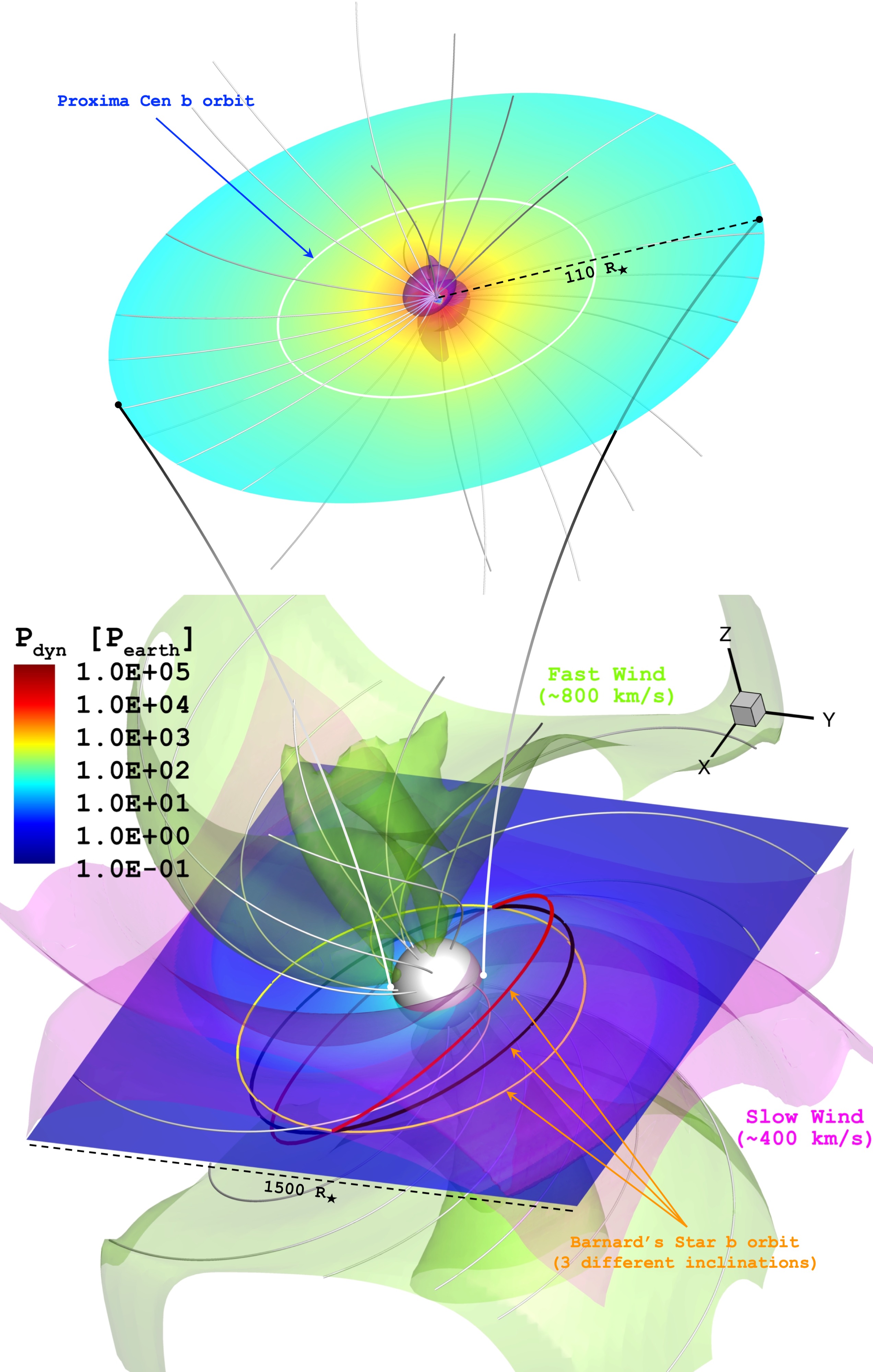}
\caption{Multi-domain numerical simulation of the stellar wind environment of Barnard's Star, driven by the ZDI map of the proxy star GJ~51 (M5V). The top panel contains the stellar corona and wind acceleration region, extending from the surface of the star up to $110$~R$_{*}$. The Alf\'en surface ($M_{\rm A} = 1$) of the stellar wind is shown in purple (see text for details). The wind solution is propagated to the extended domain presented in the bottom panel, which covers $1500$~$R_{*}$ in each cartesian direction and is centered in the star. Green and magenta iso-surfaces delineate the boundaries of the resulting fast ($U_{\rm r} \gtrsim 800$ km/s) and slow ($U_{\rm r} \lesssim 400$ km/s) wind sectors. The color scale displays the wind dynamical pressure ($P_{\rm dyn} = \rho\,U^2$), projected on the equatorial plane of both domains. The ellipses correspond to the orbit of BSb for three different inclination angles ($0, 15,$ and $30\,\deg$). The top panel also shows the orbit of Proxima Cen b for  reference. Selected magnetic field lines are shown in white.}
\label{fig_2}
\end{figure*}

The analysis of \citet{Vidotto.etal:14} also indicates objects within the range of $Ro$ estimated for Barnard's Star, are expected to have an unsigned large-scale field between $1 - 10$~G. For this reason, we have scaled the surface field strengths of all three different proxies to this range (Fig.~\ref{fig_1}). Such comparatively weak surface field is consistent with the very low level of activity measured for Barnard's Star \citep{2018arXiv181206712T}.

Finally, we stress that there are no ZDI maps for slowly-rotating fully-convective M~dwarfs that could be used as even more suitable proxies of Barnard's Star. By considering three different cases, we aim to cover observationally-motivated possibilities in field geometry, and explore how these translate to the environment of the system. In this context, the results presented here should be interpreted more as a global assessment of the range of variability of the stellar wind than absolute predictions for the system.  

\section{Results and Discussion} \label{sec:results}

\noindent Figure \ref{fig_2} presents a composite visualization of the 3D stellar wind solution achieved for Barnard's Star, using the realistically scaled surface magnetic field map of GJ~51 (Fig.~\ref{fig_1}, left panel). Both the SC and IH domains are shown, illustrating the way they are coupled in our simulations. We have performed similar wind simulations using the scaled ZDI maps of the other two proxy stars, HD~73256 and HD~179949 (Fig.~\ref{fig_1}, middle and right panels). As part of the characterization of the stellar wind solution, we include the resulting Alfv\'en surface, computed as the collection of points at which the Alfv\'enic Mach number\footnote{Defined as $M_{\rm A} = U\sqrt{4\pi\rho}/B$, where $U$, $\rho$ and $B$, are the local values of the wind speed, density, and magnetic field strength, respectively.} is equal to one. The average size of the AS is similar in all cases, ranging between $8.2$~R$_{*}$ and $10.8$ R$_{*}$, placing BSb well within the super-Alfv\'enic regime of the stellar wind (as is the case for all the solar system planets). 

\subsection{Mass Loss Rate}

\noindent Before assessing the conditions experienced by the exoplanet, we first examine the results of the simulations in the context of stellar winds from cool main sequence stars. The steady-state solutions reach maximum radial wind speeds (in km/s) of $863$ (proxy 1), $800$ (proxy 2), and $758$ (proxy 3). The respective mass loss rates in each case, expressed in solar units\footnote{$\dot{\rm M}_{\odot} \simeq 2 \times 10^{-14}$~M$_{\odot}$ yr$^{-1}$ $= 1.265 \times 10^{12}$ g s$^{-1}$} ($\dot{\rm M}_{\odot}$), are $0.085$, $0.082$, and $0.054$. 

In the absence of any observational constraints for the mass loss rate 
from Barnard's Star's wind, 
we can use the $\dot{\rm M}$ constraints for other M dwarfs to compare with the values from our numerical models. \citet{2001ApJ...547L..49W} reported an upper limit of $\dot{\rm M} < 0.2~\dot{\rm M}_{\odot}$ for Prox Cen, based on the astrospheric absorption signature appearing in the blue wing of the Lyman-$\alpha$ line. An independent $3\sigma$ upper limit of $\dot{\rm M} < 14~\dot{\rm M}_{\odot}$ ($\sim$\,$3 \times 10^{-13}$~M$_{\odot}$ yr$^{-1}$) was obtained for this star by \citet{2002ApJ...578..503W}, using a more direct technique based on the X-ray signature resulting from stellar wind ion charge exchange with ISM neutrals. A recent work focused on GJ~436 (M2.5, Age: $7 - 11$ Gyr; \citealt{2005A&A...443..609S}) suggests a relatively weak stellar wind, with $\dot{\rm M} = 0.059^{+0.074}_{-0.040}$~$\dot{\rm M}_{\odot}$, using constraints from in-transit Lyman-$\alpha$ absorption due to an evaporating outflow from the exoplanet in this system 
(see \citealt{2017MNRAS.470.4026V}).

\begin{figure}[t] 
\includegraphics[trim=6.0cm 4.6cm 9.0cm 4.25cm, clip=true,width=0.48\textwidth]{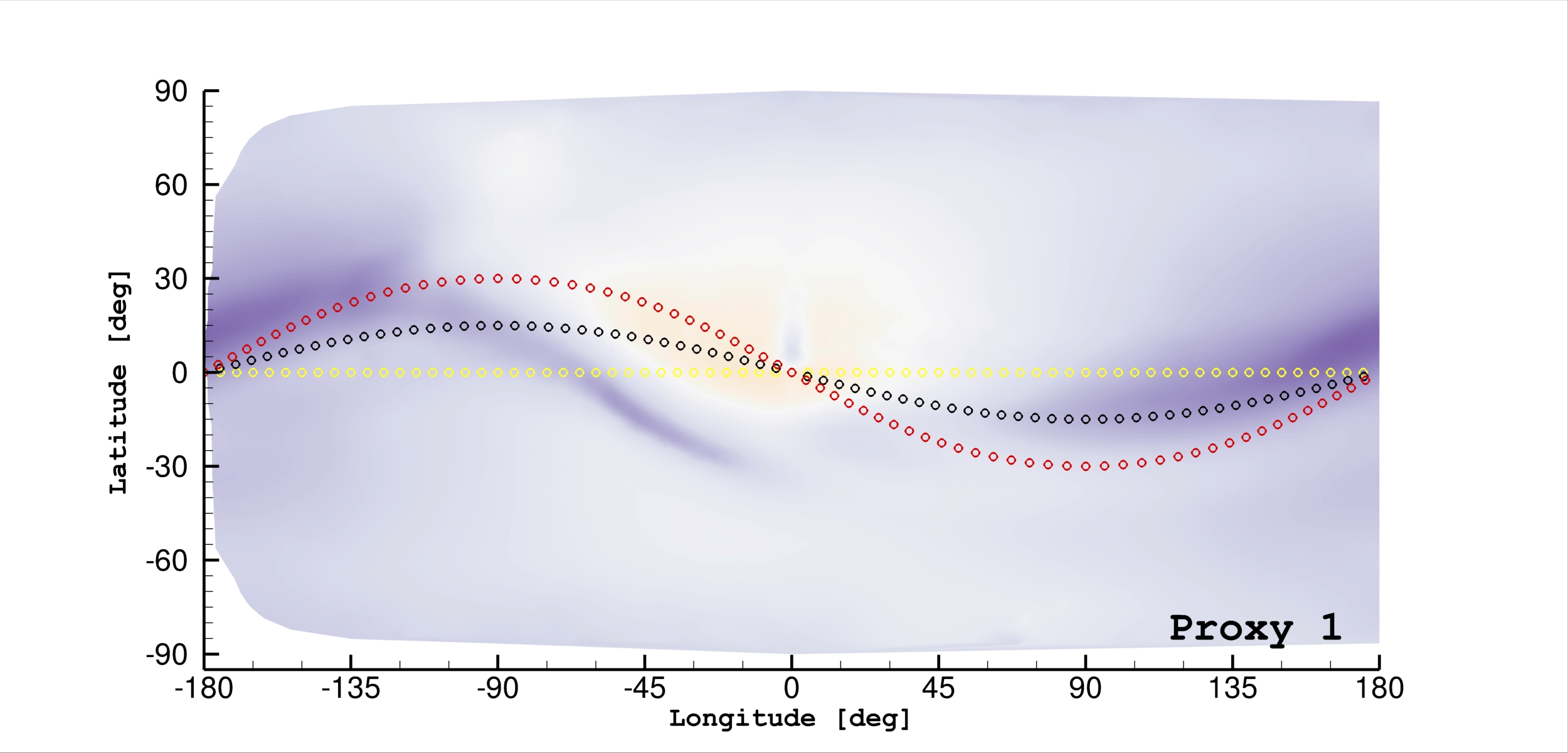}
\includegraphics[trim=6.0cm 4.6cm 9.0cm 4.25cm, clip=true,width=0.48\textwidth]{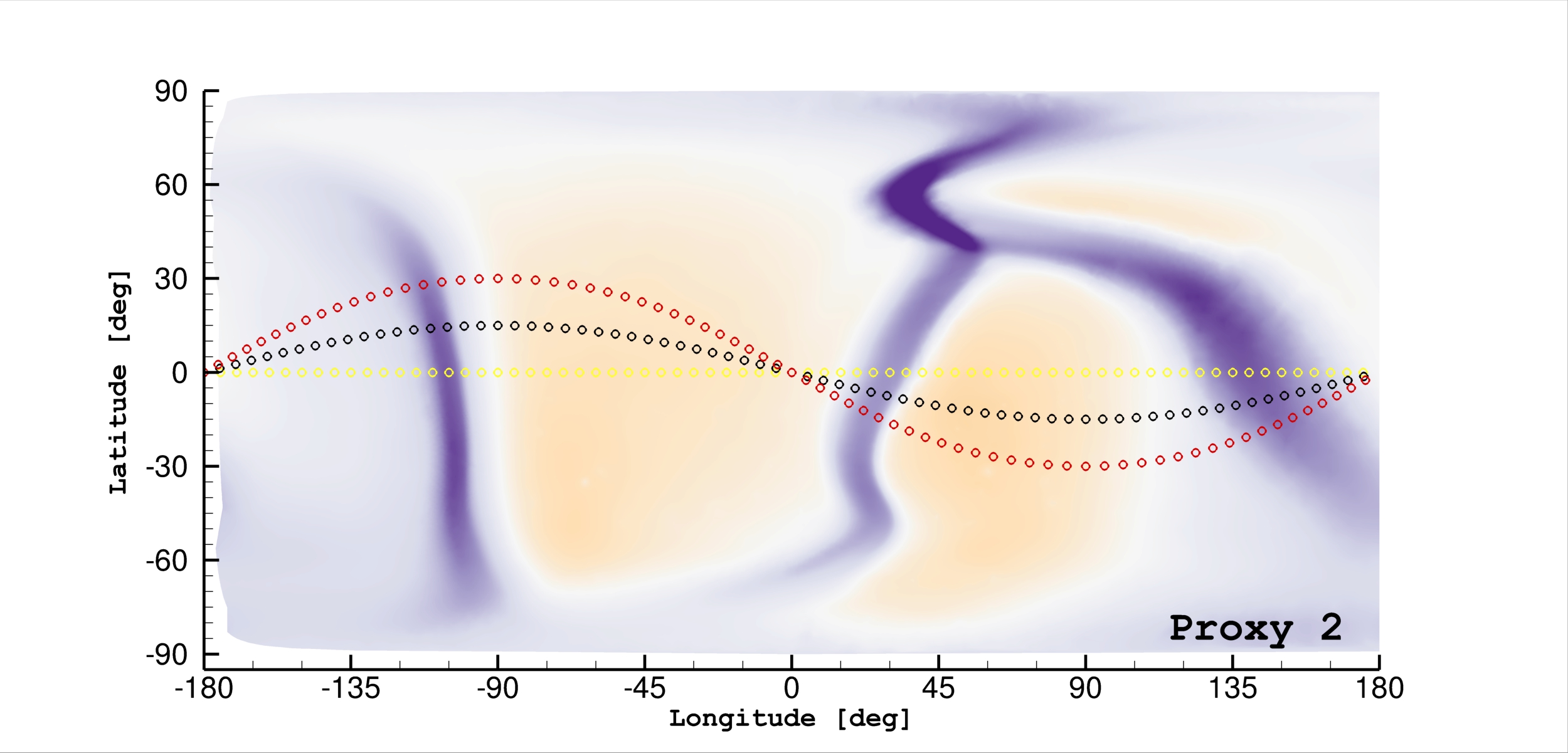}
\includegraphics[trim=6.0cm 1.0cm 9.0cm 4.25cm, clip=true,width=0.48\textwidth]{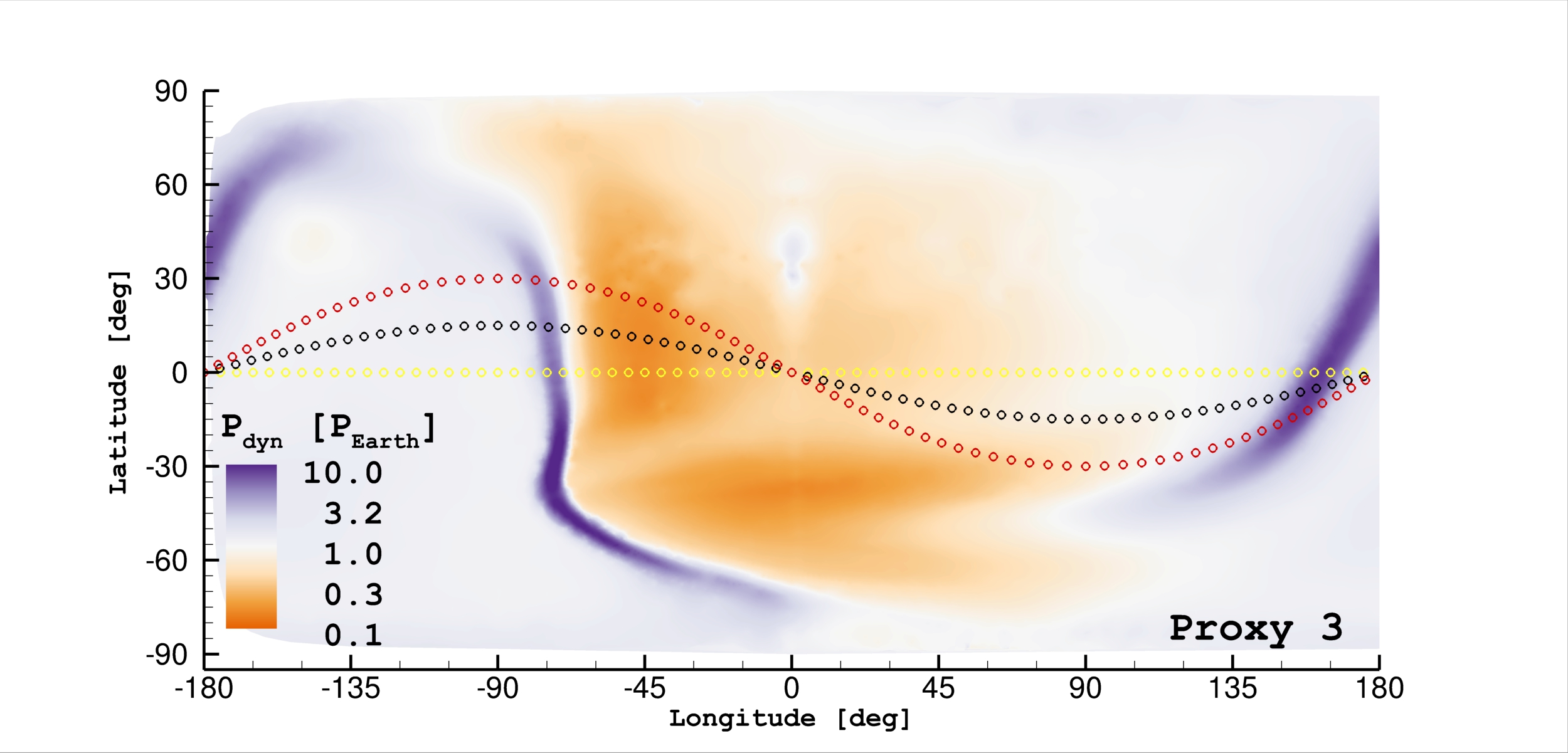}
\caption{Mercator projection of the stellar wind dynamic pressure $P_{\rm dyn}$, normalized to the solar wind pressure at 1 AU, extracted from a spheroid containing possible orbits for BSb. The different panels show the results for each proxy simulation as indicated. Dotted lines in yellow, black, and red correspond to the $0$, $15$, and $30$ deg inclination orbits, respectively (see Fig.~\ref{fig_2}, bottom panel).}
\label{fig:Pdyn_2D}
\end{figure}

\begin{figure*}{}
\center
\includegraphics[trim = 0.cm 0.cm 0.cm 0.cm,clip=true, width = 0.48
\textwidth]{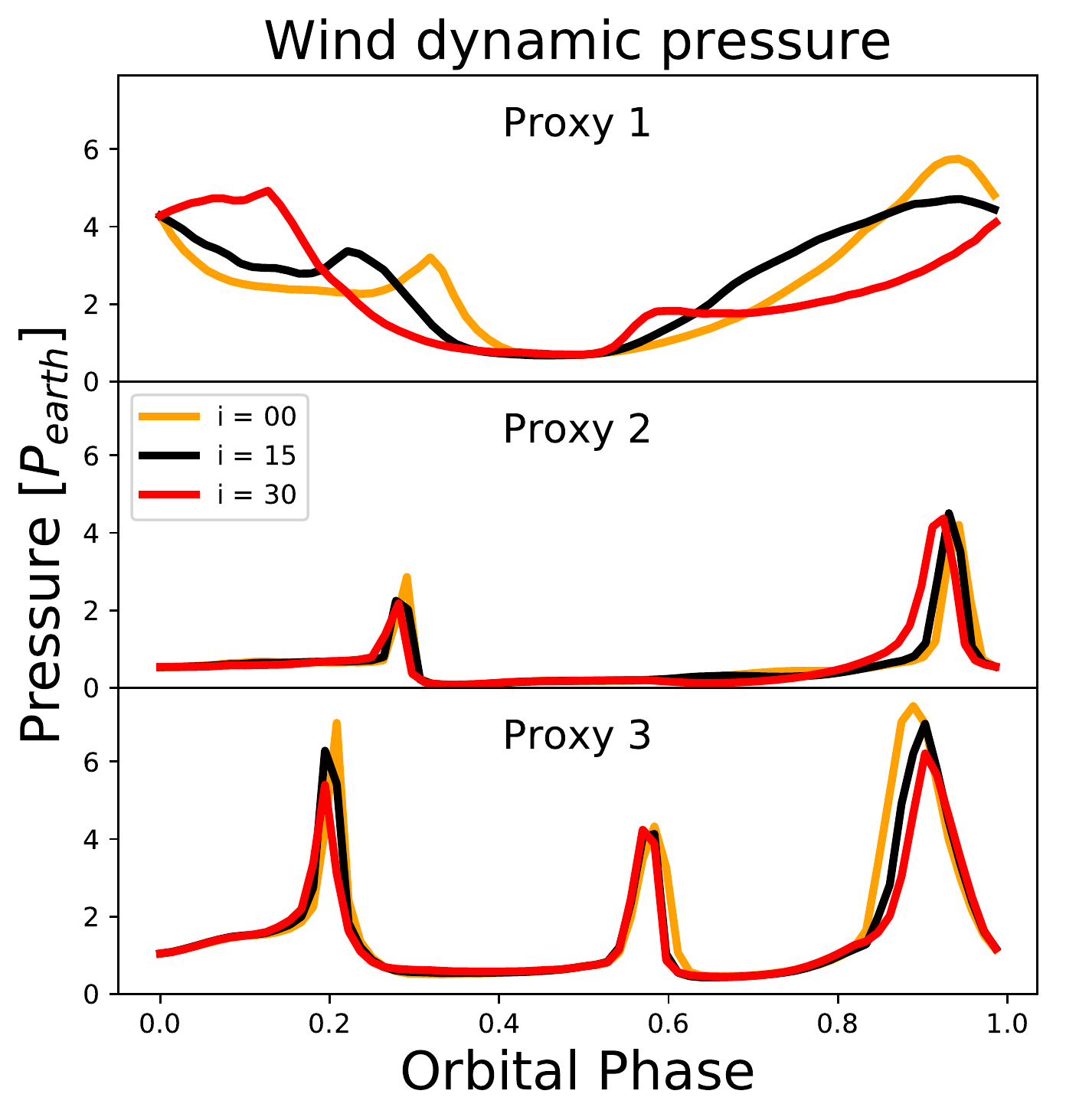}
   \includegraphics[trim = 0.cm 0.cm 0.cm 0.cm,clip=true, width =  0.48\textwidth]{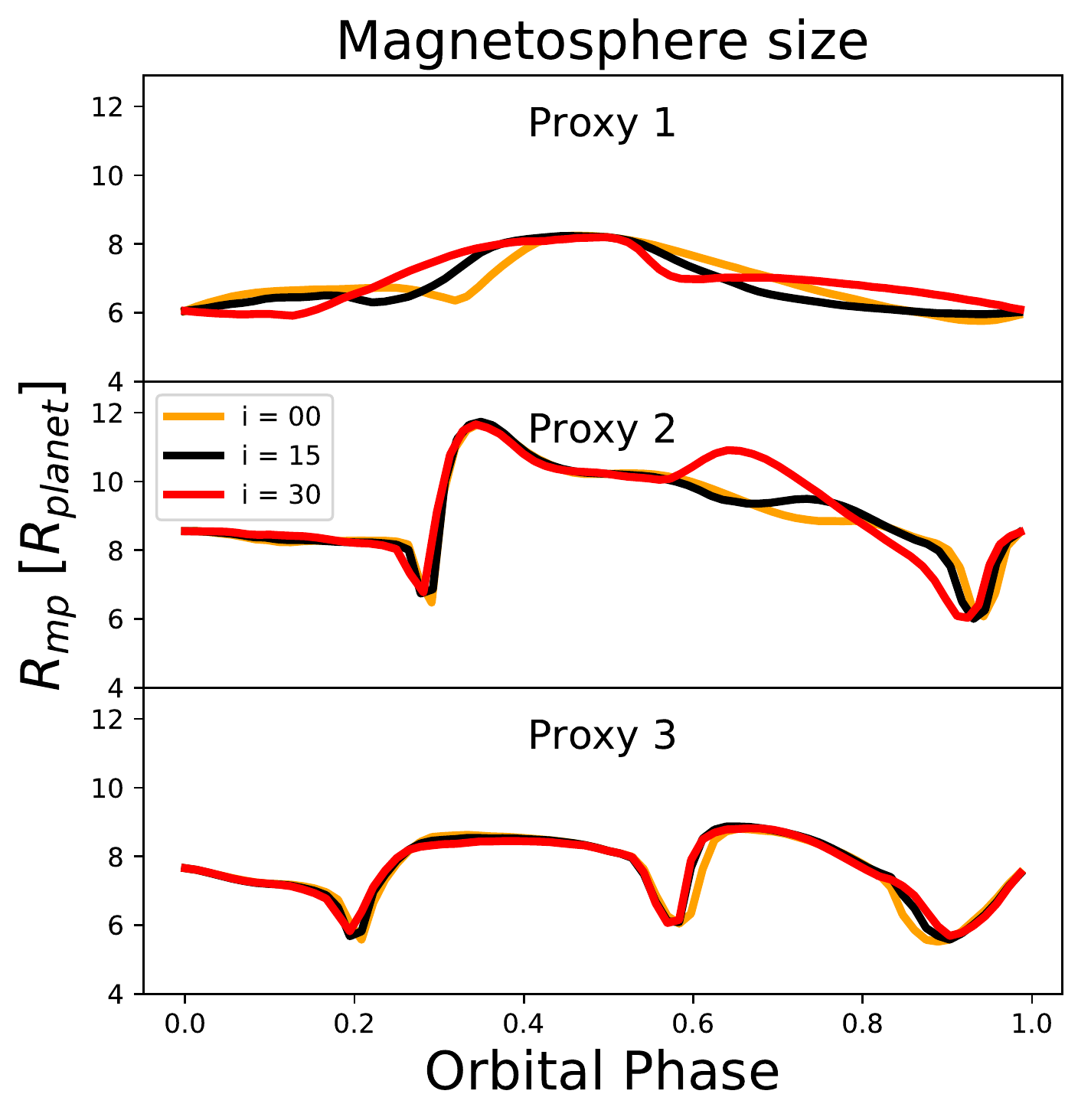}
 \caption{Behavior of $P_{\rm dyn}$ (left) and the magnetosphere size $R_{\rm mp}$ (right) along possible BSb orbits for each inclination in all of our proxies. The current sheet crossings appear as peaks (dips) in $P_{\rm dyn}$ ($R_{\rm mp}$).}
\label{fig:orbits}
\end{figure*}{}

Given the expected properties of its surface magnetic field (see~Sect.~\ref{sec:Dynamo}), Barnard's Star should in principle have an $\dot{\rm M}$ value lower\footnote{Here we are implicitly assuming that $\dot{\rm M}$ scales proportionally to the activity/magnetic flux. While the winds of Sun-like stars seem to behave in this way (up to a certain level), this assumption is very uncertain in the case of M dwarf stars due to the lack of measurements (see \citealt{2018JPhCS1100a2028W}).} than Prox Cen, and similar to to GJ 436 (provided the similarities in spectral type and age). As our numerical models are consistent with both expectations, we consider that they provide a realistic representation of the stellar wind conditions around Barnard's Star.
 
\subsection{Wind Conditions at Barnard's Star b}

\noindent The semi-major axis and eccentricity of BSb are well constrained 
while its orbital inclination is not. As can be seen in Fig.~\ref{fig_2}, the stellar wind conditions vary quite strongly within the domain. The same occurs in the remaining two proxy simulations. To quantify this, we extract the wind dynamic pressure from a spheroid containing orbits at $0$, $15$, and $30$ deg inclination angles (analogous to the ones shown in Fig.~\ref{fig_1}, bottom panel), and construct a 2D latitude-longitude Mercator projection for each proxy 
Fig.~\ref{fig:Pdyn_2D}.
$P_{\rm dyn}$ can vary up to a factor of $100$ (between $0.1$ and $10.0$ times the stellar wind conditions at Earth) for certain orbital inclinations. 
Changing the relative orientation of the orbit (i.e., the longitude of the ascending node) results in a different geometry of the 2D $P_{\rm dyn}$ projections, but their dynamic range is very similar.

In Fig.~\ref{fig:Pdyn_2D} we also include the orbital paths for the three inclinations considered. We extract the wind pressure conditions along these orbits (see Fig~\ref{fig:orbits}, left panel) and use them to compute the size of a planetary magnetosphere as in \citet{2016ApJ...833L...4G}, assuming an Earth-like planetary magnetic field of $\sim$\,$0.3$~G (right panel in Fig.~\ref{fig:orbits}) and pressure equilibrium between the stellar wind and the planet's magnetic field \citep[e.g.,][]{Schield:69,Gombosi:04} 
\begin{equation}
R_{\rm mp}/R_{\rm planet}=[B_{\rm p}^2/(8\pi P_{\rm SW})]^{1/6}.
\end{equation}
Here, $R_{mp}$ is the radius of the magnetopause, $R_{\rm planet}$ is the radius of the planet, $B_{\rm p}$ refers to the planet's equatorial magnetic field strength, and $P_{\rm SW}=n_{\rm SW}\cdot m_p\cdot U^2_{\rm SW}$, is the ram pressure of the stellar wind. 

For reference, the ram pressure of the ambient solar wind near Earth is typically about $1$ nano Pascal (nPa). During the impact of a Coronal Mass Ejection (CME), the number density lies in the range $10-50$~cm$^{-3}$, with speeds in the range of $500-2000$ km~s$^{-1}$, which translate to $1-5$~nPa for moderate CMEs, to $20-200$~nPa for the most extreme events \citep{2015GeoRL..42.4694L}.

In the case of BSb, the stellar wind pressure along the explored orbits, and for the considered proxy cases, ranges between $\sim$\,$0.1$ and $\sim$\,$7$ times the solar wind pressure at 1~AU. Assuming BSb has an Earth-like magnetic field, these wind conditions translate to a magnetospheric stand-off radius of $\sim$\,$6-12~R_{\rm planet}$, compared to the Earth's long-term average of $\sim$\,$10~R_{\rm Earth}$ \citep{2007LRSP....4....1P}.

As can be seen from Figs.~\ref{fig:Pdyn_2D} and \ref{fig:orbits}, the conditions for the proxy 1 case are on average harsher than in the other two cases, but at the same time they change more slowly (in orbital phase). This situation translates directly to the magnetosphere size, which displays the least amount of variation along the orbits (Fig.~\ref{fig:orbits}, right panel). The peaks of high pressure correspond to crossings of the current sheet, where the wind density is much higher. All orbits cross the current sheet, regardless of their inclination. Therefore, any orbit will be exposed to the higher pressures. It is good to note here that any given ZDI map case is only a snapshot of the field evolution that can occur on timescales of several days (active regions) to years (possible cycles). Therefore, the conditions inferred by our numerical models are intended to represent mean values for a given system.

In addition to modeling three possible realistic scenarios for BSb, we have also simulated the wind for a magnetic field 10 times stronger in the proxy 1 case (GJ 51), which corresponds to $20$\,\% of that used for Proxima~b in \cite{2016ApJ...833L...4G}. The results are similar in terms of geometry, differing over the domain by $\sim$\,$40\%$ in the wind speed and by roughly one order of magnitude in $\dot{\rm M}$ and $P_{\rm dyn}$. The magnetosphere size is reduced by $\sim$\,$30$\%.

This is expected to some extent: the scaling of the absolute value of the surface magnetic field plays a secondary role influencing the space weather conditions far from the star at the orbit of BSb. Rather, the important factor is the wind density declining approximately as the inverse square of the orbital distance. Stronger magnetic flux will result in just slightly faster and more massive winds (cf.  \citealt{2015ApJ...807L...6G, Reville.etal:15a, 2016A&A...594A..95A}). 

As Barnard's Star is old, its wind will have been stronger in the past. The relative importance of such conditions on the planetary atmospheric evolution of BSb would be determined by the formation mechanism that took place in this system (i.e., migration or in-situ). The inverse square scaling of wind density means a close-in planet will be surrounded by comparatively much more dense and fast varying plasma. This is what makes the expected space weather conditions on Proxima~b and TRAPPIST-1 so dramatic. Not surprisingly, BSb, at $0.4$~AU, experiences much more mild conditions than close-in planets, and comparable to days of bad space weather at Earth. 

The prognosis for atmospheric survival on BSb is much brighter than predicted for Proxima~b and the TRAPPIST-1 system.


\acknowledgments
\noindent We thank the referee for helpful suggestions. JDAG and SM were supported by Chandra grants AR4-15000X and GO5-16021X. JJD was funded by NASA contract NAS8-03060 to the CXC and thanks the Director, Belinda Wilkes, for continuing advice and support. OC was supported by NASA NExSS grant NNX15AE05G. This work used SWMF/BATSRUS tools developed at The University of Michigan Center for Space Environment Modeling. Simulations were performed on NASA's Pleiades cluster under award SMD-17-1330, provided by the NASA High-End Computing Program through the NASA Advanced Supercomputing Division at Ames Research Center.


%

\facilities{NASA Pleiades Supercomputer}


\software{SWMF \citep{2018LRSP...15....4G}}

\end{document}